\documentclass[12pt]{article}

\begin{document}

\title{A simple solution to color confinement}
\author{Johan Hansson\thanks{Part of this work was done
while visiting the \textit{Theory Group,
Lawrence Berkeley National Laboratory,
Berkeley, CA 94720, USA}. The visit was supported by
the \textit{Foundation BLANCEFLOR Boncompagni-Ludovisi, 
n\'{e}e Bildt.}} \\
\textit{Department of Physics} \\
\textit{University of G\"{a}vle} \\ 
\textit{SE-801 76 G\"{a}vle} \\ 
\textit{Sweden}}
\date{}
\maketitle

\begin{abstract}
We show that color confinement is a direct result of the nonabelian,
{\it i.e.} nonlinear, nature of the color interaction in quantum
chromodynamics. This makes it in general impossible to describe
the color field as a collection of elementary quanta (gluons).
A quark cannot be an elementary quanta of the quark field,
as the color field
of which it is the source is itself a source hence making isolated
(noninteracting) quarks impossible. 
In geometrical language, the impossibility of quarks and gluons
as physical particles arises due to the fact that the color
Yang-Mills space does not have a constant trivial curvature.
\end{abstract} 

%%%\newpage

One major problem in contemporary particle physics is
to explain why quarks and gluons are never seen as isolated
particles. A lot of effort has gone into trying to resolve this
puzzle over a period of years, including lattice QCD, dual Meissner
effect, instantons, etc. For a review, see \cite{Bander}.

We will take a different route than normally used, to
eliminate the problem before it arises.

Usually, most particle physicists use ``fields'' and ``particles''
interchangeably,
{\it i.e.} as denoting the same things. That is because the almost
universal use of Feynman diagrams gives the false impression
that particles (quanta)  are always exchanged, even when they
do not exist. The use of Feynman diagrams can be justified in
mildly nonlinear theories (weak coupling limit) but breaks down
for strongly coupled nonabelian theories. (And also for strongly
coupled abelian theories with {\it sources}.)
In quantum chromodynamics (QCD) it is at first sight a puzzle
why the color force should be short-range, and especially
 why gluons are not
seen as free particles, as the nonbroken SU(3) color symmetry
{\it seems} to demand massless quanta, which naively would have
infinite reach. However, as we shall see, there {\it are} no
quanta.

In quantum field theory a {\it particle} \cite{BjorkenDrell},
{\it i.e.} a {\it quantum} of a field,
is defined through the creation and annihilation operators, $a^{\dagger}$
and $a$.
For instance, in quantum electrodynamics (QED), the entire electromagnetic
field can be seen as a collection of superposed quanta, each
with an energy $\omega_k$. The hamiltonian of the electromagnetic field
(omitting the zero-point energy)
can be written

\begin{equation}
H  =  \sum_k N_k \omega_k, 
\end{equation}
where

\begin{equation}
 N_k = a_k^{\dagger} a_k , 
\end{equation}
is the ``number operator'', {\it i.e.} giving the number of quanta
with a specific four-momentum $k$  when operating on a (free) state,

\begin{equation}
 N_k |... n_k ...> = n_k |... n_k...>, 
\end{equation}
and, because all oscillators are independent,

\begin{equation}
 |... n_k ...> = \prod_k  | n_k>, 
\end{equation}
where $n_k$ is a positive integer, the number of quanta with that
particular momentum. The energy in the electromagnetic field
is thus the eigenvalue of the
hamiltonian (1). The reasoning for fermion fields is the same,
but then the number of quanta in the same state can be only 0 or 1.

Assuming that QCD is the true theory of quark
interactions, a problem arises, as it is generally not possible
to write the color fields in terms of superposed harmonic oscillators.
It is not possible to make a Fourier expansion and then interpret
the Fourier coefficients as creation/annihilation operators, as the
color vector potentials $A_{\mu} ^b$ ($b \in 1,...,8$), even without a 
quark current,
are governed by nonlinear evolution equations,

\begin{equation}
D^{\mu} F_{\mu \nu} = j_{\nu},
\end{equation} 
with quark current $j_{\nu} \equiv 
g_s \bar{\psi} \gamma_{\nu} \psi = 0$, we get,
in component form

\begin{equation}
(\delta_{ab} \partial^{\mu} + g_s f_{abc} A_c^{\mu})
(\partial_{\mu} A_{\nu}^b - \partial_{\nu} A_{\mu}^b + g_s f_{bde} 
A_{\mu}^d A_{\nu}^e) = 0,
\end{equation}
where $g_s$ is the color coupling constant (summation over repeated
indices implied).

When we have an abelian dynamical group, as in QED, all the structure
constants $f_{abc}$  are zero, and a general solution to Eq.(6) can be obtained
by making the Fourier expansion

\begin{equation}
A_{\mu}^{QED} = \int d^3 k \sum_{\lambda = 0}^3 a_k (\lambda) \epsilon_{\mu}(k, \lambda)
e^{-i k \cdot x} + a_k^{\dagger} (\lambda) \epsilon_{\mu}^{\ast} (k, \lambda)
e^{i k \cdot x} ,
\end{equation}
where $\epsilon_{\mu}$ is the polarization vector.

However, for a theory based on a 
 nonabelian group, like QCD, this is no longer true, due
to the nonlinear nature of Eq.(6) when $f_{abc}, f_{bde} \neq 0$,

\begin{equation}
A_{\mu}^{b \, QCD} \neq   \int d^3 k 
\sum_{\lambda = 0}^3 a_k^b (\lambda) \epsilon_{\mu}(k, \lambda)
e^{-i k \cdot x} + a_k^{b \, \dagger} (\lambda) \epsilon_{\mu}^{\ast} (k, \lambda)
e^{i k \cdot x} .
\end{equation}

The color fields can be represented by harmonic oscillators
only in the trivial, and physically empty, limit when the strong 
interaction coupling
``constant'' tends to zero, $g_s \rightarrow 0$ (or equivalently
when $Q^2 \rightarrow \infty$ because of asymptotic freedom).
Hence, no elementary quanta of the color interaction can exist.
This means that no gluon {\it particles} are possible, and that
Eq.(1) does {\it not} hold for color fields. (As another
elementary example, there is nothing within QCD which resembles
the photo-electric effect in QED, {\it i.e.} no ``gluo-electric''
effect!). The {\it fields} are always there, but their {\it quanta},
gluons and quarks,
are relevant only when probed at sufficiently (infinitely) short distances.
Generally, quarks do {\it not} exchange gluons, but the fermion fields
$q$ react to the color fields given by $A_{\mu}$. Fields are primary
to particles. 

So far we have only banished gluons. To also banish quarks as physical
particles we note that a quark field is the source of a color field,
but this color field is itself a source of a color field. Hence, a quark
field is never removed from other sources, is always interacting,
and can never be considered to be freely propagating, resulting in
that it can never be represented by harmonic oscillator modes.
This means that no quark field quanta (quarks) can ever exist.

In QED things are very different. An electric charge gives rise to 
an abelian field, which is {\it not} the source of another field.
Hence an electrically charged field {\it can} be removed from other
sources and thus exist as a physical particle. Thus, the observability
of, {\it e.g.} an electron is ultimately due to the fact that that
electromagnetic quanta (photons) can exist as real particles.

A more mathematical treatment of the physical picture given above
is provided by geometry. 
A case analogous to the one we are
studying appears in quantum field theory
on a curved spacetime \cite{BirrellDavies}, where it is well known
that fields are
more fundamental than particles.
Indeed, there it can be shown
that the very concept of a particle is, in general, useless \cite{Davies2}.
Actually nonabelian gauge fields and quantum field theory on a curved
background have a lot in common.
The total curvature, and also the dynamical coupling to
``matter fields'' through the covariant derivative,
is given by one part coming from the Yang-Mills 
connection ({\it i.e.} gauge potential)
and one part coming from the Riemannian (Levi-Civita)
connection \cite{NashSen}.
Only when {\it both} the gauge field curvature {\it and} the spacetime
curvature \cite{BirrellDavies} are zero, or at most constant,
can a particle be unambiguously
defined. The former is constant for
abelian
quantum field theory, the latter is zero on a Minkowski background with 
inertial observers, and constant for some special (and static) spacetimes.
The curvature in gauge space is given by the field strength tensor,

\begin{equation}
F_{\mu \nu}^b  =\partial_{\mu} A_{\nu}^b - \partial_{\nu} A_{\mu}^b + g_s f_{bcd} 
A_{\mu}^c A_{\nu}^d .
\end{equation}
This is the analog in gauge space to the Riemann curvature tensor ($R_{\mu \nu
\sigma \rho}$) for spacetime.
The properties of $F_{\mu \nu}$ under a gauge transformation, $U$, is
\begin{equation}
F_{\mu \nu} \rightarrow F_{\mu \nu}' = U F_{\mu \nu} U^{-1},
\end{equation}
{\it i.e.} the gauge curvature generally transforms as a tensor
in gauge space. (It is also a tensor in flat
spacetime, but not in a general Riemannian spacetime.)

However, we see directly that for an abelian gauge theory, like QED,
\begin{equation}
F_{\mu \nu} \rightarrow F_{\mu \nu}' =  F_{\mu \nu},
\end{equation}
as $U$ commutes with $F_{\mu \nu}$.
This means that the curvature is {\it constant} (invariant) in gauge space for 
an abelian field, {\it i.e.} that $F_{\mu \nu}$ transforms as a scalar
in gauge space. It also means that the field is a gauge singlet, which
only reflects that it has no ``charge'' and that the fields have no
self-interactions (abelian gauge fields $\neq$ sources of fields).
%% With just one
%%component, $F_{\mu \nu}$ can only be the scalar curvature in gauge
%%space for abelian theories.

For nonabelian fields, like QCD, the gauge curvature, $F_{\mu \nu}$ transforms
as a tensor, {\it i.e.} is {\it co}variant, not {\it in}variant, and is thus
generally {\it different} at different points in gauge space.
The color-electric and color-magnetic
fields, $\mathbf{E}^b$ and $\mathbf{B}^b$,
which are the components of $F_{\mu \nu}^b$ defined by
$F_{0i} ^ b = E_i ^b$ and $F_{ij} ^ b = \epsilon_{ijk} B_k ^b$
($i,j,k \in 1,2,3$) ,
are thus not gauge
independent and cannot be observable physical fields, which is another
way of seeing that gluons cannot exist, regardless of coupling strength.
(In contrast to usual electric and magnetic fields which {\it are}
both gauge singlets and observed.) 
Thus, color confinement is just a special case of the more general
requirement that observables be gauge invariant, {\it i.e.}
independent of the local choice of gauge ``coordinates''.
(Physically, $ F_{\mu \nu}^b \neq$ color 
singlet, implies that color gauge fields are {\it sources}
of color gauge fields.) 

We see that the unbroken nonabelian gauge theories of gravity and
QCD are strictly incompatible with the concept of (``charged'')
 particles. 
In practice, however, this only rules out gluons and quarks 
as physical particles,
as spacetime curvature (or, equivalently, observer accelerations) is
normally completely negligible in experimental settings in particle 
physics. The difference can be traced to the fact that the 
dynamical curvature is 
directly related to the (nonlinear) coupling strength, which is
enormously much larger for QCD than for gravity. Leptons can exist as 
physical particles as QED has abelian gauge dynamics and weak (nonabelian)
SU(2) is broken, {\it i.e.} absent from the point of view of particle
detectors.

It also follows, as a direct corollary of the argument above, that hadrons
must be color singlets ({\it i.e.} color neutral) as they otherwise
could not exist as physical particles.
It would be interesting to continue the analogy with gravity and speculate
that the hadrons are ``grey holes'' (as the color stays inside).
The curvature induced by the color fields would then give the hadron
(or confinement) radius.
 This would require solutions to the coupled
$q$-$A$ system (with fully dynamical quark fields), which is a very hard,
and unsolved problem.
(Strictly, also gravity should be included, perhaps in a 
Kaluza-Klein fashion, the lagrangian
then containing both $F_{\mu \nu} F^{\mu \nu}$, now with covariant
spacetime derivatives, and
$R = R^{\mu}_{\mu}$, the Ricci-scalar.)  Although this is a nice
picture, which may/may not be true, it is not necessary for the purpose 
of excluding
quarks and gluons as physical particles, for that the argument given in
this article
is sufficient.
\newpage
\noindent
In conclusion, what we have done is to provide a ``Gordian knot''-type of 
solution to color/quark confinement.
\newline
We assume only that:

1) QCD is the correct theory of quark-field interactions

2) particles (quanta) are represented by $a$ and $a^{\dagger}$
  \newline
which leads directly to the result that QCD can have {\it no} elementary quanta (gluons).
If a specific fundamental quantum
does not exist within a certain (supposedly correct) theory, it neither can be
detected in experiments. As the quark fields always generate color fields,
which in turn act as sources of other color fields, the quark fields can never be
considered to be noninteracting. Hence no expansion in harmonic oscillator
modes is possible, which means that no quark field quanta (quarks)
can exist. Only if I) QCD is wrong, or 
II) quanta are {\it not} necessarily described by harmonic oscillator
modes (which would mean that Einstein's relation $E = \hbar \omega$ does
not hold), or both, can gluons and quarks exist as physical particles.
In geometrical terms the
curvature of Yang-Mills (color) space makes quarks (as particles)
impossible. This proves that the theory of QCD automatically forbids
particles with color charge, hence implying color (gluon/quark) 
``confinement''.
In a way, there is nothing to confine in terms of particles.

\end{document}